\documentclass[%
 reprint,
 amsmath,amssymb,
 aps,
floatfix,
]{revtex4-2}

\usepackage{graphicx}
\usepackage{dcolumn}
\usepackage{bm}

\begin{document}

\preprint{APS/123-QED}

\title{Quantum size effects, multiple Dirac cones and edge states in ultrathin Bi(110) films}
\author{Asish K. Kundu}
\email{akundu@bnl.gov}
\author{Genda Gu}
\author{Tonica Valla}
\email{valla@bnl.gov}
\affiliation{Condensed Matter Physics and Materials Science Department, Brookhaven National Laboratory, Upton, New York 11973, USA}
\date{\today}

\begin{abstract}
  The presence of inherently strong spin-orbit coupling in bismuth, its unique layer-dependent band topology and high carrier mobility make it an interesting system for both fundamental studies and applications. Theoretically, it has been suggested that strong quantum size effects should be present in the Bi(110) films, with the possibility of Dirac fermion states in the odd-bilayer (BL) films, originating from dangling $p_z$ orbitals and quantum-spin hall (QSH) states in the even-bilayer films. However, the experimental verification of these claims has been lacking. Here, we study the electronic structure of Bi(110) films grown on a high-$T_c$ superconductor, Bi$_2$Sr$_2$CaCu$_2$O$_{8+\delta}$ (Bi2212) using angle-resolved photoemission spectroscopy (ARPES). We observe an oscillatory behavior of electronic structure with the film thickness and identify the Dirac-states in the odd-bilayer films, consistent with the theoretical predictions. In the even-bilayer films, we find another Dirac state that was predicted to play a crucial role in the QSH effect. In the low thickness limit, we observe several extremely one-dimensional states, probably originating from the edge-states of Bi(110) islands. Our results provide a much needed experimental insight into the electronic and structural properties of Bi(110) films.

{\bf keywords:} Ultrathin films, Electronic structure, Dirac cone, Edge states, Spin-orbit coupling, Quantum size effects, Angle-resolved photoemission spectroscopy
\end{abstract}

\maketitle

\section{Introduction}
In recent years, low-dimensional systems have attracted considerable scientific interest as they often show anomalous atomic and electronic structure and can display many exotic properties such as layer-dependent nontrivial band topology, the appearance of Dirac fermions, QSH effects, {\it etc.} which are often absent in their bulk form \cite{novoselov2004electric,li2017tunability,hofmann2006surfaces,matusalem2017deposition,galeotti2020synthesis,PhysRevB.90.195409,holtgrewe2020,bian2019survey,wang2017two,li2017effect}. In particular, group V elements (e.g., Bi and Sb) are known to show rich allotropic transformation when grown in the form of thin-films due to their semimetallic bonding character \cite{hirayama2020nucleation,nagase2018structure,nagao2004nanofilm,holtgrewe2020}. Both Bi and Sb host spin-polarized surface states, potentially applicable in spin-sources and/or filters in the field of spintronics \cite{hofmann2006surfaces}.

Depending on the substrate and the film thickness, the film surface and the associated electronic properties can be very different \cite{nagase2018structure,PhysRevB.90.195409,nagao2004nanofilm,takayama2015one,peng2019unusual,yamada2018ultrathin}. For example, bulk bismuth crystalizes in a rhombohedral A7 structure (space group R-3m) with two atoms per unit cell. In bismuth films, the hexagonal (HEX) surface, corresponding to Bi(111) is the preferred structure for thicker films, while the pseudocubic (PC) phase, corresponding to the (110) surface is more stable for thinner films \cite{PhysRevB.80.245407,shimamura2018ultrathin,yamada2018ultrathin}. According to theoretical studies, both PC and HEX phases should show exotic surface electronic structure with Dirac states in the interior and edge-states at step edges, that could be topological in some circumstances \cite{PhysRevB.90.195409,li2017tunability,bian2019survey,takayama2015one,drozdov2014one,kim2014edge}. The electronic character of odd-bilayer Bi(110) films is expected to be very different from Bi(111) films since one nearest neighbor bond of Bi is broken, leaving a dangling bond \cite{PhysRevB.77.045428,PhysRevB.90.195409}. In the even-bilayer Bi(110) films, the dangling bonds are removed due to the inter-bilayer covalent bonding \cite{PhysRevB.90.195409}. Theoretical studies have predicted ample of interesting properties in Bi(110) films. For example, the Dirac states originating from the surface dangling bonds at $\bar{M}$ point of surface Brillouin zone (SBZ) are predicted in odd-bilayer films \cite{PhysRevB.90.195409} while QSH states should exist in the even-bilayer films below a critical film thickness of $\leq$ 4 BL. The Dirac-states near $E_F$ could lead to unconventional optical and transport properties, such as enhanced spin-polarized transport, while the QSH could be used in dissipationless transport devices.

Topological edge-states, were predicted \cite{PhysRevLett.97.236805} and experimentally detected at  Bi(111) surfaces using scanning tunneling microscopy/spectroscopy (STM/STS) \cite{drozdov2014one,peng2018visualizing,kim2014edge}. Later, an ARPES study questioned the topological nature of these states \cite{takayama2015one}. Topological edge-states have also been observed in Bi(110) films with a black-phosphorus (BP)-like structure (no buckling within a bilayer) grown on highly-oriented pyrolytic graphite (HOPG), and Nb(110) substrates \cite{lu2015topological,yang2017edge,gou2020effect}. Such topological states were not observed in Bi(110) films with a distorted BP-like structure (buckled bilayers) grown on Si(111) \cite{nagao2004nanofilm,nagase2018structure} and 4H-SiC(0001) \cite{sun2012energy}. Theoretical studies suggest that atomic buckling plays a crucial role in determining the topological character of these films \cite{lu2015topological,li2017tunability}. Below the critical value of buckling, $h= 0.1$ {\AA} and the film thickness ($\leq$ 4 BL), the Bi(110) films should be nontrivial and above that, they should be trivial \cite{lu2015topological}. However, the dispersion of the Bi(110) edge-states has never been experimentally observed. Furthermore, the surface structure of Bi(110) films, a crucial parameter that determines the outcome of theoretical studies, is still debated.
\begin{figure*}[ht]
\centering
\includegraphics[width=10cm]{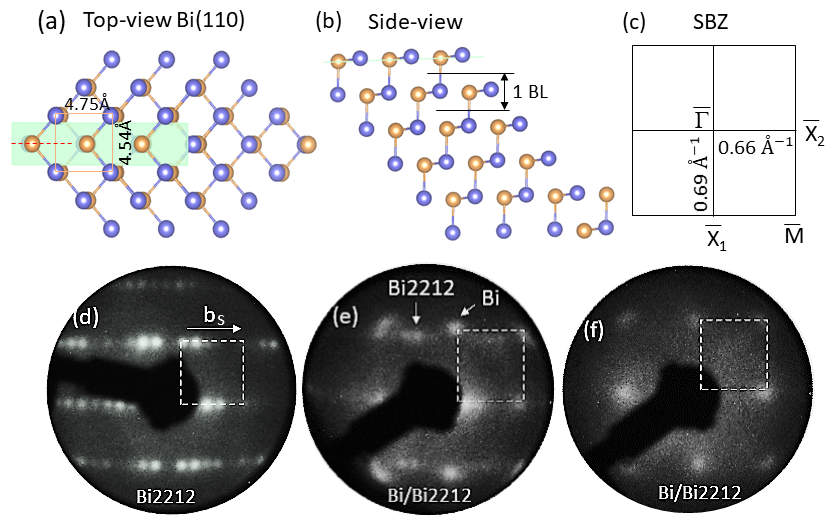}
\caption {Truncated A7-bulk structure of Bi(110) and the LEED pattern of substrate (Bi2212) as well as Bi films. (a) Top view of Bi(110) where pseudo-square unit cell is marked. Dotted line represent the mirror plane. (b) Side view (parallel to the mirror plane). Within 1 BL (3.28 {\AA}) two consecutive atoms (different colors) share different atomic position, resulting in buckled structure. (c) Schematic of the Bi(110) surface Brillouin zone (SBZ). (d) LEED pattern of substrate (crystallographic $\vec{b_s}$ -axis is marked), (e) 3 BL, and (f) 5 BL Bi films on Bi2212, respectively. Dotted square in (d) and (e, f) represent the reciprocal unit cell of Bi2212 and Bi(110) films, respectively.}\label{Fig1}
\end{figure*}

The Bi(110) films have been synthesized on various substrates  \cite{nagao2004nanofilm,PhysRevB.80.245407,yaginuma2007electronic,kokubo2015ultrathin,nagase2018structure,PhysRevB.90.195409,kowalczyk2020realization,hatta2009epitaxial,sun2012energy,shimamura2018ultrathin,yamada2018ultrathin}. It has been found that on HOPG, only the even bilayer islands nucleate, whereas on Si(111)$(\surd3\times\surd3)$-B and Bi2212 both the even and odd bilayer Bi(110) islands coexist. The experimental studies were mostly performed by using STM, with some ARPES results, typically lacking a systematic thickness dependence  \cite{PhysRevB.80.245407,PhysRevB.90.195409,bian2019survey,kowalczyk2020realization,yamada2018ultrathin}. Therefore, very important questions such as the existence of Dirac-bands in the odd-bilayer Bi(110) films, or gapped Dirac states necessary for the edge-states, as well as the dispersion of the latter in the even-bilayer films remain open.

To answer some of these questions, we have grown Bi(110) films {\it in-situ} on optimally doped Bi2212 substrate and studied their layer-dependent structural and electronic properties using low-energy electron diffraction (LEED) and ARPES. Our results show that both the even- and odd bilayer islands grow on Bi2212. For the first time, we provide direct evidence for the Dirac cone at the $\bar{M}$ point of SBZ in the odd-bilayer Bi(110) films. We also observe another Dirac state of different origins in the odd-bilayer Bi(110) films. Finally, in the low-thickness limit, we observe a highly 1D-like state with the dispersion that resembles the theoretically predicted dispersion of the edge-states. Our results also provide evidence for different surface structures of even and odd bilayer films.

\section{Results and discussions}
\subsection{Surface structure of Bi(110) films}

The top and side views of Bi(110) surface with an A7 crystal structure are shown in figure~\ref{Fig1}(a) and (b), respectively. The pseudocubic surface unit cell is marked by a rectangle in figure~\ref{Fig1}(a), with the corresponding SBZ shown in figure~\ref{Fig1}(c). The cleaved Bi2212 surface shows a typical diffraction pattern with the long-range supermodulation along the $b_S$ direction, figure~\ref{Fig1}(d) \cite{lindberg1988surface}. The LEED pattern of a film with the nominal thickness of 3 BL, figure~\ref{Fig1}(e), displays both the substrate and overlayer-Bi related spots, an indication that films do not grow in a layer-by-layer mode, but follow a three-dimensional (3D) island growth. The similar growth mode was also observed for other substrates \cite{hatta2009epitaxial,nagase2018structure, shimamura2018ultrathin}. The observed pseudo-square LEED pattern indicates the formation of Bi(110) films. As the lattice parameters of Bi(110) surface ($a=4.54$ {\AA}, $b=4.75$ {\AA}) are smaller than the lattice parameter of Bi2212 ($a_S\approx b_S\approx5.4$ {\AA}), the Bi(110) reciprocal unit-cell is bigger than the Bi2212 one. The arc-like feature of the second-order diffraction spots is the signature of coexisting orthogonal domains originating from different lattice parameters, $a$ and $b$. Similar LEED pattern was reported in previous studies of Bi/Bi2212 system \cite{shimamura2018ultrathin}. The LEED patterns indicate that the Bi-O-plane of the Bi2212 substrate and Bi(110) films are aligned so that $a\parallel b_S$ and $b\parallel b_S$.

At a nominal coverage of $\sim$ 5 BL, the substrate spots are no longer visible, figure~\ref{Fig1}(f). Considering the atomic structure of Bi(110) surface, the islands are expected to propagate along the zigzag chain as Bi atoms are only bonded along that direction, whereas the atoms in adjacent chains are not bonded directly in plane. Therefore, during the growth, migrating atoms tend to attach to the ends of the zigzag chains. STM studies on Bi(110) indicate that the longer side of islands is the zigzag direction ({\it i.e.}, the shorter side of the unit-cell) \cite{nagase2018structure}. If this is also true for the Bi/Bi2212 system, the obtained LEED pattern (figure~\ref{Fig1}(e)) implies that the Bi islands grow so that the $\bar{\Gamma}-\bar{X_1}$ and $\bar{\Gamma}-\bar{X_2}$ directions are along the $a_S$ and $b_S$ crystallographic axes of Bi2212. Indeed, in the previous STM study, it was apparent that islands nucleated along the two specific orthogonal directions \cite{shimamura2018ultrathin}. However, these studies could not resolve whether the islands had BP or A7 structures, as both structures have very similar unit cells.

\subsection{Electronic structure of Bi(110) films}

\begin{figure*}[ht!]
\centering
\includegraphics[width=17cm]{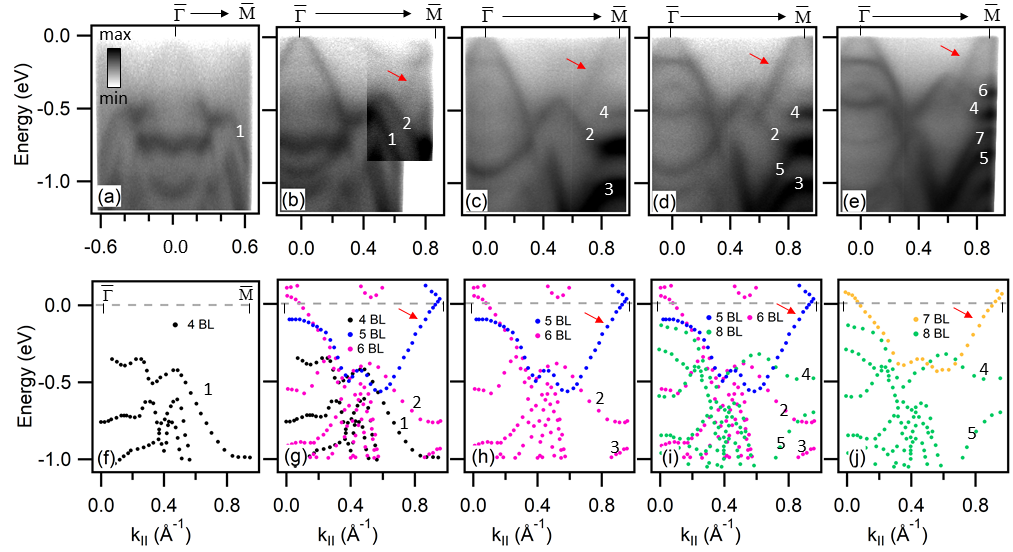}
\caption{Measured (top panels) and theoretical (bottom panels) electronic structure of Bi(110) films along the $\bar{\Gamma}$-$\bar{M}$ for various film thicknesses. (a)-(e) Experimental band dispersions for the nominal film thickness of 2 BL, 3 BL, 3.5 BL, 4 BL, and 5 BL, respectively. (f)-(j) Calculated layer-dependent bands of various bilayers as indicated. Bands are reproduced from Ref. \protect\cite{lu2015topological} (even-bilayer) and Ref. \protect\cite{bian2019survey} (odd-bilayer), respectively. To match the experimental data, a rigid downward shift of 130 (90) meV was applied to the calculated bands of 4 (8) BL films. Experimental bands marked by numbers and arrows in (a)-(e) can be identified in theoretical calculations (f)-(j). Bands indicated by arrows are the Dirac states originating from the dangling bonds. The bands 6 and 7 probably originate from the 10 BL islands.}\label{Fig2}
\end{figure*}

The experimental band dispersions along the $\bar{\Gamma}-\bar{M}$ direction of Bi(110) SBZ for the nominal film coverage of $\sim2$ BL, $\sim3$ BL, $\sim3.5$ BL , $\sim4$ BL and $\sim5$ BL are presented in figure~\ref{Fig2}(a)-(e). With the increasing Bi coverage, a complex evolution of electronic bands indicates that multiple film thicknesses are present simultaneously, in accordance with our LEED results. To understand the origin of individual bands, the theoretical layer-dependent dispersions are reproduced from Ref. \cite{lu2015topological} (even-bilayer) and Ref. \cite{bian2019survey} (odd-bilayer), figure~\ref{Fig2}(f)-(j). The comparison enables the identification of spectroscopic features corresponding to the specific film thickness. For example, in the case of even-bilayer films, going from 4 BL to higher coverage, the electronic states marked by 1, 2, and 4 in figure~\ref{Fig2}(f)-(i) show very similar dispersions. On the other hand, the calculations predict quite a different behavior for the odd-bilayer films, with a linear dispersive band and Dirac cone at $\bar{M}$ (figure~\ref{Fig2}(g)-(j)) \cite{bian2019survey,PhysRevB.90.195409}. Variation of spectral intensity of a particular band with the nominal Bi coverage is related to the change in the relative area covered by the corresponding thickness.

According to theory, in the 2 BL and 4 BL films (figure~\ref{Fig2}(f)) there should be essentially no electronic states crossing the $E_F$ and a spectral gap should be observed \cite{lu2015topological,bian2019survey}. Our results for the nominal film coverage of $\sim$ 2 BL, figure~\ref{Fig2}(a), show prominent signatures corresponding to 4 BL Bi(110) films, with some contribution from 3 BL and/or 6 BL thickness forming a hole-like band at $\bar{\Gamma}$. The states originating from 4 BL islands show a  spectral gap along $\bar{\Gamma}-\bar{M}$, in agreement with the calculations.

The film with a total nominal thickness of $\sim$ 3 BL, shows a few more states appearing (figure~\ref{Fig2}(b)). Comparison with the theoretical result (figure~\ref{Fig2}(g)) indicates that the additional states are mainly originating from the 6 BL islands. The presence of a linear dispersive band, marked by an arrow in figure~\ref{Fig2}(b), suggests an additional contribution from odd-bilayer islands, most likely from the 3 BL and/or 5 BL islands. The exact discrimination between them is difficult due to its extremely low intensity. Thus, the 3 BL nominal thickness film consists of 4 BL, 5 BL (and/or 3 BL), and 6 BL Bi(110) islands and small regions of uncovered Bi2212, as inferred from LEED (figure~\ref{Fig1}(e)). We note that there are some additional states which are visible near $E_F$ (0 to $-$0.3 eV) in between $\bar{\Gamma}$ and $\bar{M}$. These states are absent in the layer-dependent band calculations and their intensity is strongly suppressed with further Bi deposition.

Figure~\ref{Fig2}(c) shows the electronic states for the nominal coverage of 3.5 BL film. Now, the electronic states corresponding to 4 BL are suppressed, and the film is dominated by 6 BL Bi(110) islands. The electronic states 2 and 3 are well reproduced by theoretical calculations for 6 BL Bi(110) film (figure~\ref{Fig2}(h)). In addition, the presence of a less intense band 4 indicates a small contribution from 8 BL islands. Upon further Bi deposition with a nominal coverage $\sim$ 4 BL, the bands 4 and 5, originating from 8 BL Bi(110) islands get sharper and well-resolved as can be seen in figure~\ref{Fig2}(d). The linearly dispersing band near the $\bar{M}$ point also gained in intensity compared to the 3.5 BL film. This band shows a local minimum at $\sim$~k$_\|$=0.6 {\AA$^{-1}$ and its corresponding energy position of $\sim$ $-$0.57 eV matches well with the theoretical values of 5 BL Bi(110) film (figure~\ref{Fig2}(i)). Upon further Bi deposition, the local minimum further shifts to $\sim$ $-$0.47 eV (figure~\ref{Fig2}(e)) corresponding to the 7 BL Bi(110) film (figure~\ref{Fig2}(j)). The observed trend would suggest that the bands marked by 6 and 7 in figure~\ref{Fig2}(e) originate from the 10 BL islands.

\subsection{Fermi surface of Bi(110)}

\begin{figure*}[ht!]
\centering
\includegraphics[width=14cm]{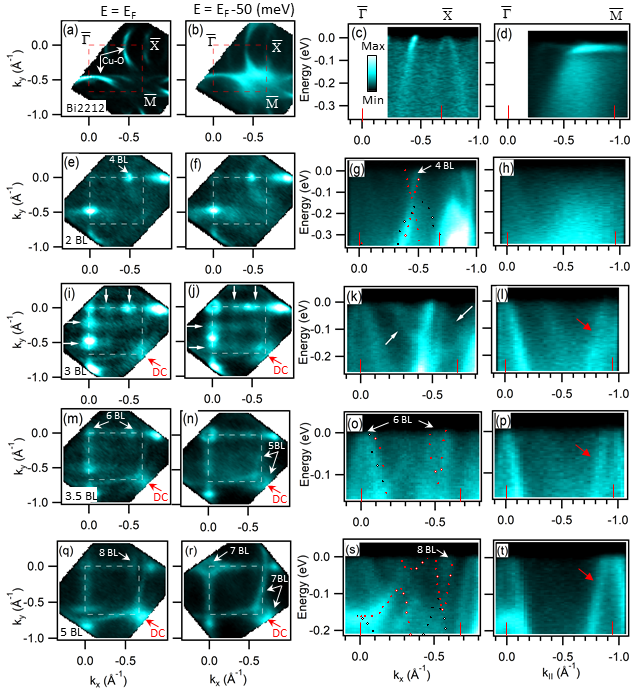}
\caption {Fermi surface and band dispersions along the high symmetry directions of Bi(110) SBZ for bare substrate and various film thickness of Bi films. (a)-(d) substrate (Bi2212), (e)-(h) 2 BL, (i)-(l) 3 BL, (m)-(p) 3.5 BL, and (q)-(t) 5 BL films, respectively. For the FS plots, the photoemission intensities are integrated within $\pm$8 meV around Fermi level. The dotted square represents the one quadrant of the Bi(110) SBZ. Dirac-cone (DC) and its corresponding linear dispersive band at $\bar{M}$ are marked by the arrows (red). Electronic states corresponding to the CuO plane of Bi2212 are indicated in (a). White color arrows in (i) and (j) indicating the 1D FS. The dispersions of these states are indicated by arrows in (k). Calculated band dispersions of 4 BL, 6 BL, and 8 BL Bi(110) films are reproduced from Ref. \cite{lu2015topological} and superimposed on (g), (o) and (s), respectively. Where the red and black dots represent calculated bands along $\bar{\Gamma}$-$\bar{X_1}$ and $\bar{\Gamma}$-$\bar{X_2}$ directions of SBZ (figure 1(c)), respectively. }\label{Fig3}
\end{figure*}
The Fermi surfaces and constant-energy contours at $-50$ meV, along with the band dispersions for bare substrate and various film thicknesses are shown in figure~\ref{Fig3}: (a-d) bare Bi2212, (e-h) 2 BL, (i-l) 3 BL, (m-p) 3.5 BL and (q-t) 5 BL, respectively. In figure~\ref{Fig3}(a), the states originating from the Cu-O plane of the bare Bi2212 substrate are marked. The Fermi surface, dispersion and superconducting gap are consistent with our previous results on optimally doped Bi2212 \cite{drozdov2018,Valla2020}. After the deposition of 2 BL film, the intensity of these substrate related states is strongly suppressed whereas an intense spot from the Bi film appears near the $\bar{X}$ point along the $\bar{\Gamma}$-$\bar{X}$ (figure~\ref{Fig3}(e) and (f)). The $k_F$ position of this spot agrees with the expected $k_F$ from the 4 BL Bi(110) islands  \cite{kowalczyk2013,lu2015topological}.

According to calculations, this band hosts a massive Dirac cone with a gap size of $\sim$ 70 meV, formed by the bonding and antibonding states of Bi-$p_z$ orbitals \cite{lu2015topological,li2017tunability}.}
In figure~\ref{Fig3}(g), the theoretical dispersions are shown on top of the experimental data. The bands for both $\bar{\Gamma}$-$\bar{X_1}$ (red dots) and $\bar{\Gamma}$-$\bar{X_2}$ (black dots) directions are shown, as they should coexist due to the formation of orthogonal domains, as discussed earlier. The red dots represent the Dirac cone.
The more detailed picture of the experimental electronic structure of that Dirac state is given in figure~\ref{Fig4}. The Dirac point is located $\sim$~150 meV below $E_F$. Such Dirac cone does not exist on the surface of bulk Bi(110) crystal \cite{hofmann2006surfaces}.
\begin{figure}[ht!]
\centering
\includegraphics[width=8.5cm]{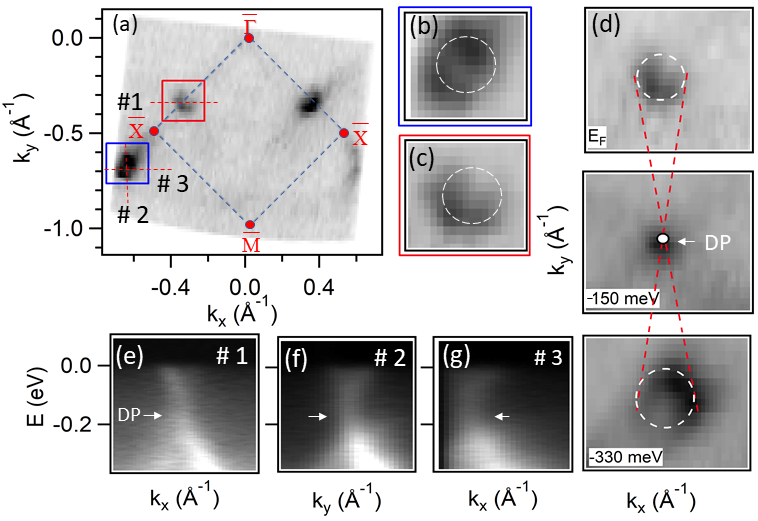}
\caption {Fermi surface and band dispersions of 2 BL Bi(110) film (nominal thickness). (a) Fermi surface with the zoomed-in views of features inside blue (b) and red (c) boxes, as indicated. (d) Variation of constant energy contours for the state inside the red box from (a, c). (e)-(g) Energy-momentum dispersions of the Dirac-like features along the momentum cuts indicated in (a). Arrows in  (d)-(e) indicate the Dirac point (DP). The photoemission intensity is given by the gray-scale, with high-intensity corresponding to black in (a)-(d) and white in (e)-(g), respectively}\label{Fig4}
\end{figure}

\begin{figure*}[ht]
\centering
\includegraphics[width=16cm]{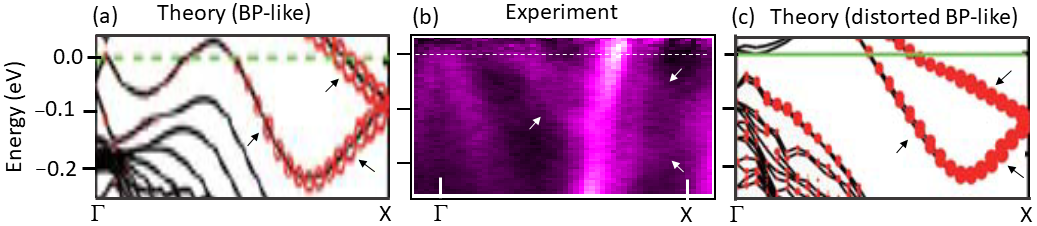}
\caption {Comparison of theoretical \protect\cite{lu2015topological} and experimental edge-state dispersions in thin 3 BL Bi(110) films. (a) Theoretical edge-state dispersion for 6 nm Bi(110) nano-ribbon with BP-like (topological) and (c) the distorted BP-like structure (trivial) \protect\cite{lu2015topological}. (b) Experimentally observed edge-state dispersion, marked with the white arrows. Arrows in (a) and (c) represents the edge-states those are identified experimentally.}\label{S3}
\end{figure*}

In the case of 3 BL coverage, the FS (figure~\ref{Fig3}(i)) looks more complex and in addition to the 4 BL, the features from 3 BL and/or 5 BL and 6 BL are also visible. In figure~\ref{Fig3}(k) and (l), the hole-like band at $\bar{\Gamma}$, predominantly originating from 6 BL islands, creates a circular feature in the FS (figure~\ref{Fig3}(i)) and in the $-50$ meV constant-energy contour (figure~\ref{Fig3}(j)). The Dirac-cone at $\bar{M}$, marked by the red arrows in figure~\ref{Fig3}(i), (j) and (l) originates from the 3 BL and/or 5 BL islands. Overall, these features are in good agreement with the calculated FS of 3 BL/5 BL \cite{PhysRevB.77.045428} and 6 BL \cite{kowalczyk2013} Bi(110) films. Interestingly, we also observe a set of new features that form extremely one-dimensional (1D) FS, running orthogonal to each other, as indicated by the white arrows in figure~\ref{Fig3}(i). These features are better visualized in the second derivative of spectral intensity, as shown in the supplementary information, figure S1. They were not observed on the Bi(110) surface of a bulk crystal \cite{agergaard2001effect} and do not appear in the layer-dependent FS calculations \cite{PhysRevB.77.045428,kowalczyk2013}. We also note that they quickly fade away for thicker films (figure~\ref{Fig3}(q) and (r)). One feasible interpretation would be that these states are the edge-states of Bi(110) islands. Their 1D-like FSs are aligned with the edges of the Bi(110) islands, $\bar{\Gamma}$-$\bar{X}$. Their energy-momentum dispersions as indicated by arrows in figure~\ref{Fig3}(k) are also in good agreement with the calculated edge-states of Bi(110) ribbons \cite{lu2015topological,li2017tunability}.
A direct side-by-side comparison of experimental and theoretical dispersion of these states is further presented in figure~\ref{S3}. The theoretical edge-state calculations for 6 nm wide Bi(110) nano-ribbon with BP-like (panel (a)) and the distorted BP-like structure (panel (c)) [19] are compared to the experimentally observed edge-state dispersion (panel (b)) of 3 BL films. The measured states resemble the edge-states from calculations. However, their relatively low-intensity and the presence of other bands make it difficult to discriminate between the BP-like and distorted BP-like structure of the film. We note that the Bi(110) edge-states were previously observed in the STM measurements \cite{lu2015topological,yang2017edge,gou2020effect}, but their full energy-momentum dispersion has never been reported before.

The 3.5 BL film (figure~\ref{Fig3}(m)-(p)), does not show any 4 BL-related features, while the 1D-like features are now suppressed. The spectra show mainly 6 BL (indicated in figure~\ref{Fig3}(m)) and 5 BL-related features (Dirac cone at $\bar{M}$).

In the case of 5 BL films, the elongated FS segments, marked by \lq\lq{}8 BL\rq\rq{} in figure~\ref{Fig3}(q) are not originating from the edge-states, but from the electronic states of 8 BL films, as shown in figure~\ref{Fig3}(s). The main differences between 3.5 BL and 5 BL films are that the circular feature (figure~\ref{Fig3}(q) and (r)) and the linearly dispersing bands (figure~\ref{Fig3}(s) and (t)) around $\bar{\Gamma}$ originate predominantly from the 7 BL islands, not from 6 BL. Furthermore, the presence of intense spots at $\bar{M}$ in figure~\ref{Fig3}(m) and (q), clearly suggest that the Dirac point in these films are very close to $E_F$, in good agreement with the theoretical results (figure~\ref{Fig2}(j)). The Dirac cone at $\bar{M}$ was also observed at the surface of bulk Bi(110) due to the presence of surface dangling bonds \cite{hofmann2006surfaces}. We also note the 3 BL, 3.5 BL and 5 BL films all show an arc-like feature along the $\bar{X_1}-\bar{M}-\bar{X_2}$ in the FS (figure~\ref{Fig3}(i), (m) and (q)) and in the constant-energy contour (figure~\ref{Fig3}(j), (n) and (r)). This feature is due to the presence of a shallow electron pocket that is also present in the theoretical calculations for 3 BL and 5 BL Bi(110) films \cite{PhysRevB.77.045428}.

Further, we want to point out that, although the ARPES is generally not sensitive to the edge-state, as the probed volume fraction of the edges is usually extremely small, in the case of Bi/Bi2212, the edge density might just be large enough to generate a detectable photoemission signal. In thinner films, the islands are relatively small, with widths much smaller than the island lengths \cite{shimamura2018ultrathin}, while in thicker films the islands generally grow bigger and the effective edge density within the ARPES probing region decreases.

As already noted, the theoretical results \cite{lu2015topological,li2017tunability} show that the 2 BL and 4 BL Bi(110) films should host gapped Dirac states along $\bar{\Gamma}-\bar{X}$ $-$ the prerequisite conditions to support QSH states \cite{li2017tunability}. Our results on 4 BL islands (figure~\ref{Fig3}(g) and figure~\ref{Fig4}) suggest that the films are metallic and without a significant gap at the Dirac point. The absence of a gap would suggest that the observed 1D-like states in our experiment might be the edge resonances, rather than in-gap edge-localized states. This would further suggest that it is unlikely that the Bi/Bi2212 system hosts QSH states. However, as the magnitude, the position relative to the Fermi level and the character of the gap are determined by the atomic buckling, lattice parameter, and charge transfer at the film-substrate interface, it is expected that by growing Bi films on appropriate substrates, the Dirac state could be tuned so that it supports a QSH effect \cite{lu2015topological,li2017tunability}. We also expect that an application of hydrostatic or uniaxial pressure could be used for fine tuning of a given system into a QHS state.

It is important that, unlike in Bi films on HOPG, where only even bilayer islands are formed, both even and odd bilayer films can be stabilized on Bi2212. This suggests that the substrate plays a crucial role in stabilizing the nucleation of islands that might be otherwise energetically unfavorable. The electronic fingerprint of the odd-bilayer islands is a Dirac-like band at $\bar{M}$ \cite{PhysRevB.90.195409}. Theoretical study shows that this cone is a consequence of the dangling $p_z$ orbitals on the surface, the strong spin-orbit coupling and mirror symmetry of the lattice \cite{PhysRevB.90.195409}. Thus, experimental observation of this feature confirms the existence of surface dangling bonds which in turn would imply the A7-like (bulk-like) surface structure of the odd-bilayer films.

A direct comparison of experimental data with the BP-like and A7-like structure can be seen in figure S2, suggesting that the surface structure of even-bilayer films is most likely very close to the BP-like. Further, considering the theoretical prediction that the Bi(110) films ($>$ 4 BL) with BP-like structure are topologically trivial and metallic \cite{lu2015topological,li2017tunability} would suggest that the edge-like states detected in our study are likely topologically trivial. The spin-polarised ARPES and STM studies would be helpful in addressing this issue.

\section{Summary}

In summary, we have studied the electronic structure of Bi(110) films for various thicknesses and discovered the multiple Dirac states in the electronic structure. The possibility of a Dirac cone for odd-bilayer films was predicted theoretically, but our study offers the first experimental confirmation. The observed Dirac bands are a consequence of mirror symmetry of the lattice and a strong spin-orbit coupling \cite{PhysRevB.90.195409}. Our results also suggest the A7-like surface structure of the odd-bilayer and the BP-like of the even-bilayer films. We also observe a highly 1D-like FS in thinner films, suggestive of the edge-states localized to the edges of small but dense (110) islands. Further studies, including the STM and spin-resolved ARPES, would be necessary to resolve whether these 1D states are topologically trivial or nontrivial. We also show that for even bilayer films, the required conditions for QSH effects are not fulfilled in the Bi/Bi2212 system and indicate that such states could be achieved by tuning film-substrate interaction. Our results provide a solid basis for a complete understanding of the electronic structure of this system and indicate a possibility of using the chiral surface states (Dirac state) of the Bi(110) films in spintronics. However, in accordance with our previous studies, an electron transfer from the Bi films renders the interface region of the Bi2212 substrate to become strongly hole-underdoped and non-superconducting \cite{kundu2020origin,yilmaz2014absence}. This effectively prohibits the  superconducting proximity effect and prevents this interesting system from becoming a platform for topological superconductivity.

\section{Experimental}
Bi films were grown on \textit{in-situ} cleaved optimally doped Bi2212 single crystals. The Bi2212 substrates were synthesized by the traveling floating zone method, clamped to the sample holder and cleaved with Kapton tape in the ARPES preparation chamber. The high purity bismuth (99.99\%) was evaporated from a resistively heated Al$_2$O$_3$ crucible. During the film growth, substrate was kept at room temperature (RT). The photoemission experiments were carried out on a Scienta SES-R4000 electron spectrometer with the monochromatized He I$_\alpha$(21.2 eV) and He II$_\alpha$(40.8 eV) radiation (VUV-5k) \cite{Kim2018a}. The total instrumental energy resolution was $\sim$ 20 meV and 30 meV for the ARPES and core-level measurements, respectively. The angular resolution was better than 0.15$^{\circ}$ and 0.3$^{\circ}$ along and perpendicular to the slit of the analyzer, respectively. Most of the data were taken at $\sim$30 K. The nominal film thickness was determined by measuring the attenuation of Bi 5$d_{5/2}$ core-levels of Bi2212 substrate \cite{kundu2020origin}.

\section*{Acknowledgements}
This work was supported by the US Department of Energy, Office of Basic Energy Sciences, contract no. DE-SC0012704.

\bibliography{refBi}

\begin{thebibliography}{39}%
\makeatletter
\providecommand \@ifxundefined [1]{%
 \@ifx{#1\undefined}
}%
\providecommand \@ifnum [1]{%
 \ifnum #1\expandafter \@firstoftwo
 \else \expandafter \@secondoftwo
 \fi
}%
\providecommand \@ifx [1]{%
 \ifx #1\expandafter \@firstoftwo
 \else \expandafter \@secondoftwo
 \fi
}%
\providecommand \natexlab [1]{#1}%
\providecommand \enquote  [1]{``#1''}%
\providecommand \bibnamefont  [1]{#1}%
\providecommand \bibfnamefont [1]{#1}%
\providecommand \citenamefont [1]{#1}%
\providecommand \href@noop [0]{\@secondoftwo}%
\providecommand \href [0]{\begingroup \@sanitize@url \@href}%
\providecommand \@href[1]{\@@startlink{#1}\@@href}%
\providecommand \@@href[1]{\endgroup#1\@@endlink}%
\providecommand \@sanitize@url [0]{\catcode `\\12\catcode `\$12\catcode
  `\&12\catcode `\#12\catcode `\^12\catcode `\_12\catcode `\%12\relax}%
\providecommand \@@startlink[1]{}%
\providecommand \@@endlink[0]{}%
\providecommand \url  [0]{\begingroup\@sanitize@url \@url }%
\providecommand \@url [1]{\endgroup\@href {#1}{\urlprefix }}%
\providecommand \urlprefix  [0]{URL }%
\providecommand \Eprint [0]{\href }%
\providecommand \doibase [0]{https://doi.org/}%
\providecommand \selectlanguage [0]{\@gobble}%
\providecommand \bibinfo  [0]{\@secondoftwo}%
\providecommand \bibfield  [0]{\@secondoftwo}%
\providecommand \translation [1]{[#1]}%
\providecommand \BibitemOpen [0]{}%
\providecommand \bibitemStop [0]{}%
\providecommand \bibitemNoStop [0]{.\EOS\space}%
\providecommand \EOS [0]{\spacefactor3000\relax}%
\providecommand \BibitemShut  [1]{\csname bibitem#1\endcsname}%
\let\auto@bib@innerbib\@empty
\bibitem [{\citenamefont {Novoselov}\ \emph {et~al.}(2004)\citenamefont
  {Novoselov}, \citenamefont {Geim}, \citenamefont {Morozov}, \citenamefont
  {Jiang}, \citenamefont {Zhang}, \citenamefont {Dubonos}, \citenamefont
  {Grigorieva},\ and\ \citenamefont {Firsov}}]{novoselov2004electric}%
  \BibitemOpen
  \bibfield  {author} {\bibinfo {author} {\bibfnamefont {K.~S.}\ \bibnamefont
  {Novoselov}}, \bibinfo {author} {\bibfnamefont {A.~K.}\ \bibnamefont {Geim}},
  \bibinfo {author} {\bibfnamefont {S.~V.}\ \bibnamefont {Morozov}}, \bibinfo
  {author} {\bibfnamefont {D.}~\bibnamefont {Jiang}}, \bibinfo {author}
  {\bibfnamefont {Y.}~\bibnamefont {Zhang}}, \bibinfo {author} {\bibfnamefont
  {S.~V.}\ \bibnamefont {Dubonos}}, \bibinfo {author} {\bibfnamefont {I.~V.}\
  \bibnamefont {Grigorieva}},\ and\ \bibinfo {author} {\bibfnamefont {A.~A.}\
  \bibnamefont {Firsov}},\ }\bibfield  {title} {\bibinfo {title} {Electric
  field effect in atomically thin carbon films},\ }\href@noop {} {\bibfield
  {journal} {\bibinfo  {journal} {Science}\ }\textbf {\bibinfo {volume}
  {306}},\ \bibinfo {pages} {666} (\bibinfo {year} {2004})}\BibitemShut
  {NoStop}%
\bibitem [{\citenamefont {Li}\ \emph {et~al.}(2017{\natexlab{a}})\citenamefont
  {Li}, \citenamefont {Ji}, \citenamefont {Li}, \citenamefont {Hu},
  \citenamefont {Cai}, \citenamefont {Zhang},\ and\ \citenamefont
  {Yan}}]{li2017tunability}%
  \BibitemOpen
  \bibfield  {author} {\bibinfo {author} {\bibfnamefont {S.-s.}\ \bibnamefont
  {Li}}, \bibinfo {author} {\bibfnamefont {W.-x.}\ \bibnamefont {Ji}}, \bibinfo
  {author} {\bibfnamefont {P.}~\bibnamefont {Li}}, \bibinfo {author}
  {\bibfnamefont {S.-j.}\ \bibnamefont {Hu}}, \bibinfo {author} {\bibfnamefont
  {L.}~\bibnamefont {Cai}}, \bibinfo {author} {\bibfnamefont {C.-w.}\
  \bibnamefont {Zhang}},\ and\ \bibinfo {author} {\bibfnamefont {S.-s.}\
  \bibnamefont {Yan}},\ }\bibfield  {title} {\bibinfo {title} {Tunability of
  the quantum spin hall effect in bi(110) films: Effects of electric field and
  strain engineering},\ }\href@noop {} {\bibfield  {journal} {\bibinfo
  {journal} {ACS Applied Materials \& Interfaces}\ }\textbf {\bibinfo {volume}
  {9}},\ \bibinfo {pages} {21515} (\bibinfo {year}
  {2017}{\natexlab{a}})}\BibitemShut {NoStop}%
\bibitem [{\citenamefont {Hofmann}(2006)}]{hofmann2006surfaces}%
  \BibitemOpen
  \bibfield  {author} {\bibinfo {author} {\bibfnamefont {P.}~\bibnamefont
  {Hofmann}},\ }\bibfield  {title} {\bibinfo {title} {The surfaces of bismuth:
  Structural and electronic properties},\ }\href@noop {} {\bibfield  {journal}
  {\bibinfo  {journal} {Progress in surface science}\ }\textbf {\bibinfo
  {volume} {81}},\ \bibinfo {pages} {191} (\bibinfo {year} {2006})}\BibitemShut
  {NoStop}%
\bibitem [{\citenamefont {Matusalem}\ \emph {et~al.}(2017)\citenamefont
  {Matusalem}, \citenamefont {Koda}, \citenamefont {Bechstedt}, \citenamefont
  {Marques},\ and\ \citenamefont {Teles}}]{matusalem2017deposition}%
  \BibitemOpen
  \bibfield  {author} {\bibinfo {author} {\bibfnamefont {F.}~\bibnamefont
  {Matusalem}}, \bibinfo {author} {\bibfnamefont {D.~S.}\ \bibnamefont {Koda}},
  \bibinfo {author} {\bibfnamefont {F.}~\bibnamefont {Bechstedt}}, \bibinfo
  {author} {\bibfnamefont {M.}~\bibnamefont {Marques}},\ and\ \bibinfo {author}
  {\bibfnamefont {L.~K.}\ \bibnamefont {Teles}},\ }\bibfield  {title} {\bibinfo
  {title} {Deposition of topological silicene, germanene and stanene on
  graphene-covered sic substrates},\ }\href@noop {} {\bibfield  {journal}
  {\bibinfo  {journal} {Scientific reports}\ }\textbf {\bibinfo {volume} {7}},\
  \bibinfo {pages} {1} (\bibinfo {year} {2017})}\BibitemShut {NoStop}%
\bibitem [{\citenamefont {Galeotti}\ \emph {et~al.}(2020)\citenamefont
  {Galeotti}, \citenamefont {De~Marchi}, \citenamefont {Hamzehpoor},
  \citenamefont {MacLean}, \citenamefont {Rao}, \citenamefont {Chen},
  \citenamefont {Besteiro}, \citenamefont {Dettmann}, \citenamefont {Ferrari},
  \citenamefont {Frezza} \emph {et~al.}}]{galeotti2020synthesis}%
  \BibitemOpen
  \bibfield  {author} {\bibinfo {author} {\bibfnamefont {G.}~\bibnamefont
  {Galeotti}}, \bibinfo {author} {\bibfnamefont {F.}~\bibnamefont {De~Marchi}},
  \bibinfo {author} {\bibfnamefont {E.}~\bibnamefont {Hamzehpoor}}, \bibinfo
  {author} {\bibfnamefont {O.}~\bibnamefont {MacLean}}, \bibinfo {author}
  {\bibfnamefont {M.~R.}\ \bibnamefont {Rao}}, \bibinfo {author} {\bibfnamefont
  {Y.}~\bibnamefont {Chen}}, \bibinfo {author} {\bibfnamefont {L.}~\bibnamefont
  {Besteiro}}, \bibinfo {author} {\bibfnamefont {D.}~\bibnamefont {Dettmann}},
  \bibinfo {author} {\bibfnamefont {L.}~\bibnamefont {Ferrari}}, \bibinfo
  {author} {\bibfnamefont {F.}~\bibnamefont {Frezza}}, \emph {et~al.},\
  }\bibfield  {title} {\bibinfo {title} {Synthesis of mesoscale ordered
  two-dimensional $\pi$-conjugated polymers with semiconducting properties},\
  }\href@noop {} {\bibfield  {journal} {\bibinfo  {journal} {Nature Materials}\
  ,\ \bibinfo {pages} {1}} (\bibinfo {year} {2020})}\BibitemShut {NoStop}%
\bibitem [{\citenamefont {Bian}\ \emph {et~al.}(2014)\citenamefont {Bian},
  \citenamefont {Wang}, \citenamefont {Miller}, \citenamefont {Chiang},
  \citenamefont {Kowalczyk}, \citenamefont {Mahapatra},\ and\ \citenamefont
  {Brown}}]{PhysRevB.90.195409}%
  \BibitemOpen
  \bibfield  {author} {\bibinfo {author} {\bibfnamefont {G.}~\bibnamefont
  {Bian}}, \bibinfo {author} {\bibfnamefont {X.}~\bibnamefont {Wang}}, \bibinfo
  {author} {\bibfnamefont {T.}~\bibnamefont {Miller}}, \bibinfo {author}
  {\bibfnamefont {T.-C.}\ \bibnamefont {Chiang}}, \bibinfo {author}
  {\bibfnamefont {P.~J.}\ \bibnamefont {Kowalczyk}}, \bibinfo {author}
  {\bibfnamefont {O.}~\bibnamefont {Mahapatra}},\ and\ \bibinfo {author}
  {\bibfnamefont {S.~A.}\ \bibnamefont {Brown}},\ }\bibfield  {title} {\bibinfo
  {title} {First-principles and spectroscopic studies of bi(110) films:
  Thickness-dependent dirac modes and property oscillations},\ }\href
  {https://doi.org/10.1103/PhysRevB.90.195409} {\bibfield  {journal} {\bibinfo
  {journal} {Phys. Rev. B}\ }\textbf {\bibinfo {volume} {90}},\ \bibinfo
  {pages} {195409} (\bibinfo {year} {2014})}\BibitemShut {NoStop}%
\bibitem [{\citenamefont {Holtgrewe}\ \emph {et~al.}(2020)\citenamefont
  {Holtgrewe}, \citenamefont {Mahatha}, \citenamefont {Sheverdyaeva},
  \citenamefont {Moras}, \citenamefont {Flammini}, \citenamefont {Colonna},
  \citenamefont {Ronci}, \citenamefont {Papagno}, \citenamefont {Barla},
  \citenamefont {Petaccia} \emph {et~al.}}]{holtgrewe2020}%
  \BibitemOpen
  \bibfield  {author} {\bibinfo {author} {\bibfnamefont {K.}~\bibnamefont
  {Holtgrewe}}, \bibinfo {author} {\bibfnamefont {S.}~\bibnamefont {Mahatha}},
  \bibinfo {author} {\bibfnamefont {P.}~\bibnamefont {Sheverdyaeva}}, \bibinfo
  {author} {\bibfnamefont {P.}~\bibnamefont {Moras}}, \bibinfo {author}
  {\bibfnamefont {R.}~\bibnamefont {Flammini}}, \bibinfo {author}
  {\bibfnamefont {S.}~\bibnamefont {Colonna}}, \bibinfo {author} {\bibfnamefont
  {F.}~\bibnamefont {Ronci}}, \bibinfo {author} {\bibfnamefont
  {M.}~\bibnamefont {Papagno}}, \bibinfo {author} {\bibfnamefont
  {A.}~\bibnamefont {Barla}}, \bibinfo {author} {\bibfnamefont
  {L.}~\bibnamefont {Petaccia}}, \emph {et~al.},\ }\bibfield  {title} {\bibinfo
  {title} {Topologization of $\beta$-antimonene on bi 2 se 3 via proximity
  effects},\ }\href@noop {} {\bibfield  {journal} {\bibinfo  {journal}
  {Scientific reports}\ }\textbf {\bibinfo {volume} {10}},\ \bibinfo {pages}
  {1} (\bibinfo {year} {2020})}\BibitemShut {NoStop}%
\bibitem [{\citenamefont {Bian}\ \emph {et~al.}(2019)\citenamefont {Bian},
  \citenamefont {Wang}, \citenamefont {Kowalczyk}, \citenamefont {Maerkl},
  \citenamefont {Brown},\ and\ \citenamefont {Chiang}}]{bian2019survey}%
  \BibitemOpen
  \bibfield  {author} {\bibinfo {author} {\bibfnamefont {G.}~\bibnamefont
  {Bian}}, \bibinfo {author} {\bibfnamefont {X.}~\bibnamefont {Wang}}, \bibinfo
  {author} {\bibfnamefont {P.~J.}\ \bibnamefont {Kowalczyk}}, \bibinfo {author}
  {\bibfnamefont {T.}~\bibnamefont {Maerkl}}, \bibinfo {author} {\bibfnamefont
  {S.~A.}\ \bibnamefont {Brown}},\ and\ \bibinfo {author} {\bibfnamefont
  {T.-C.}\ \bibnamefont {Chiang}},\ }\bibfield  {title} {\bibinfo {title}
  {Survey of electronic structure of bi and sb thin films by first-principles
  calculations and photoemission measurements},\ }\href@noop {} {\bibfield
  {journal} {\bibinfo  {journal} {Journal of Physics and Chemistry of Solids}\
  }\textbf {\bibinfo {volume} {128}},\ \bibinfo {pages} {109} (\bibinfo {year}
  {2019})}\BibitemShut {NoStop}%
\bibitem [{\citenamefont {Wang}\ \emph {et~al.}(2017)\citenamefont {Wang},
  \citenamefont {Ji}, \citenamefont {Zhang}, \citenamefont {Li}, \citenamefont
  {Zhang}, \citenamefont {Wang}, \citenamefont {Li},\ and\ \citenamefont
  {Yan}}]{wang2017two}%
  \BibitemOpen
  \bibfield  {author} {\bibinfo {author} {\bibfnamefont {Y.-p.}\ \bibnamefont
  {Wang}}, \bibinfo {author} {\bibfnamefont {W.-x.}\ \bibnamefont {Ji}},
  \bibinfo {author} {\bibfnamefont {C.-w.}\ \bibnamefont {Zhang}}, \bibinfo
  {author} {\bibfnamefont {P.}~\bibnamefont {Li}}, \bibinfo {author}
  {\bibfnamefont {S.-f.}\ \bibnamefont {Zhang}}, \bibinfo {author}
  {\bibfnamefont {P.-j.}\ \bibnamefont {Wang}}, \bibinfo {author}
  {\bibfnamefont {S.-s.}\ \bibnamefont {Li}},\ and\ \bibinfo {author}
  {\bibfnamefont {S.-s.}\ \bibnamefont {Yan}},\ }\bibfield  {title} {\bibinfo
  {title} {Two-dimensional arsenene oxide: A realistic large-gap quantum spin
  hall insulator},\ }\href@noop {} {\bibfield  {journal} {\bibinfo  {journal}
  {Applied Physics Letters}\ }\textbf {\bibinfo {volume} {110}},\ \bibinfo
  {pages} {213101} (\bibinfo {year} {2017})}\BibitemShut {NoStop}%
\bibitem [{\citenamefont {Li}\ \emph {et~al.}(2017{\natexlab{b}})\citenamefont
  {Li}, \citenamefont {Ji}, \citenamefont {Hu}, \citenamefont {Zhang},\ and\
  \citenamefont {Yan}}]{li2017effect}%
  \BibitemOpen
  \bibfield  {author} {\bibinfo {author} {\bibfnamefont {S.-s.}\ \bibnamefont
  {Li}}, \bibinfo {author} {\bibfnamefont {W.-x.}\ \bibnamefont {Ji}}, \bibinfo
  {author} {\bibfnamefont {S.-j.}\ \bibnamefont {Hu}}, \bibinfo {author}
  {\bibfnamefont {C.-w.}\ \bibnamefont {Zhang}},\ and\ \bibinfo {author}
  {\bibfnamefont {S.-s.}\ \bibnamefont {Yan}},\ }\bibfield  {title} {\bibinfo
  {title} {Effect of amidogen functionalization on quantum spin hall effect in
  bi/sb (111) films},\ }\href@noop {} {\bibfield  {journal} {\bibinfo
  {journal} {ACS applied materials \& interfaces}\ }\textbf {\bibinfo {volume}
  {9}},\ \bibinfo {pages} {41443} (\bibinfo {year}
  {2017}{\natexlab{b}})}\BibitemShut {NoStop}%
\bibitem [{\citenamefont {Hirayama}(2020)}]{hirayama2020nucleation}%
  \BibitemOpen
  \bibfield  {author} {\bibinfo {author} {\bibfnamefont {H.}~\bibnamefont
  {Hirayama}},\ }\bibfield  {title} {\bibinfo {title} {Nucleation and growth of
  ultrathin bi films},\ }\href@noop {} {\bibfield  {journal} {\bibinfo
  {journal} {Advances in Physics: X}\ }\textbf {\bibinfo {volume} {6}},\
  \bibinfo {pages} {1845975} (\bibinfo {year} {2020})}\BibitemShut {NoStop}%
\bibitem [{\citenamefont {Nagase}\ \emph {et~al.}(2018)\citenamefont {Nagase},
  \citenamefont {Kokubo}, \citenamefont {Yamazaki}, \citenamefont {Nakatsuji},\
  and\ \citenamefont {Hirayama}}]{nagase2018structure}%
  \BibitemOpen
  \bibfield  {author} {\bibinfo {author} {\bibfnamefont {K.}~\bibnamefont
  {Nagase}}, \bibinfo {author} {\bibfnamefont {I.}~\bibnamefont {Kokubo}},
  \bibinfo {author} {\bibfnamefont {S.}~\bibnamefont {Yamazaki}}, \bibinfo
  {author} {\bibfnamefont {K.}~\bibnamefont {Nakatsuji}},\ and\ \bibinfo
  {author} {\bibfnamefont {H.}~\bibnamefont {Hirayama}},\ }\bibfield  {title}
  {\bibinfo {title} {Structure and growth of bi(110) islands on si(111)
  3$\times$ 3-b substrates},\ }\href@noop {} {\bibfield  {journal} {\bibinfo
  {journal} {Physical Review B}\ }\textbf {\bibinfo {volume} {97}},\ \bibinfo
  {pages} {195418} (\bibinfo {year} {2018})}\BibitemShut {NoStop}%
\bibitem [{\citenamefont {Nagao}\ \emph {et~al.}(2004)\citenamefont {Nagao},
  \citenamefont {Sadowski}, \citenamefont {Saito}, \citenamefont {Yaginuma},
  \citenamefont {Fujikawa}, \citenamefont {Kogure}, \citenamefont {Ohno},
  \citenamefont {Hasegawa}, \citenamefont {Hasegawa},\ and\ \citenamefont
  {Sakurai}}]{nagao2004nanofilm}%
  \BibitemOpen
  \bibfield  {author} {\bibinfo {author} {\bibfnamefont {T.}~\bibnamefont
  {Nagao}}, \bibinfo {author} {\bibfnamefont {J.}~\bibnamefont {Sadowski}},
  \bibinfo {author} {\bibfnamefont {M.}~\bibnamefont {Saito}}, \bibinfo
  {author} {\bibfnamefont {S.}~\bibnamefont {Yaginuma}}, \bibinfo {author}
  {\bibfnamefont {Y.}~\bibnamefont {Fujikawa}}, \bibinfo {author}
  {\bibfnamefont {T.}~\bibnamefont {Kogure}}, \bibinfo {author} {\bibfnamefont
  {T.}~\bibnamefont {Ohno}}, \bibinfo {author} {\bibfnamefont {Y.}~\bibnamefont
  {Hasegawa}}, \bibinfo {author} {\bibfnamefont {S.}~\bibnamefont {Hasegawa}},\
  and\ \bibinfo {author} {\bibfnamefont {T.}~\bibnamefont {Sakurai}},\
  }\bibfield  {title} {\bibinfo {title} {Nanofilm allotrope and phase
  transformation of ultrathin bi film on si(111)-7$\times$ 7},\ }\href@noop {}
  {\bibfield  {journal} {\bibinfo  {journal} {Phys. Rev. Lett.}\ }\textbf
  {\bibinfo {volume} {93}},\ \bibinfo {pages} {105501} (\bibinfo {year}
  {2004})}\BibitemShut {NoStop}%
\bibitem [{\citenamefont {Takayama}\ \emph {et~al.}(2015)\citenamefont
  {Takayama}, \citenamefont {Sato}, \citenamefont {Souma}, \citenamefont
  {Oguchi},\ and\ \citenamefont {Takahashi}}]{takayama2015one}%
  \BibitemOpen
  \bibfield  {author} {\bibinfo {author} {\bibfnamefont {A.}~\bibnamefont
  {Takayama}}, \bibinfo {author} {\bibfnamefont {T.}~\bibnamefont {Sato}},
  \bibinfo {author} {\bibfnamefont {S.}~\bibnamefont {Souma}}, \bibinfo
  {author} {\bibfnamefont {T.}~\bibnamefont {Oguchi}},\ and\ \bibinfo {author}
  {\bibfnamefont {T.}~\bibnamefont {Takahashi}},\ }\bibfield  {title} {\bibinfo
  {title} {One-dimensional edge states with giant spin splitting in a bismuth
  thin film},\ }\href@noop {} {\bibfield  {journal} {\bibinfo  {journal} {Phys.
  Rev. Lett.}\ }\textbf {\bibinfo {volume} {114}},\ \bibinfo {pages} {066402}
  (\bibinfo {year} {2015})}\BibitemShut {NoStop}%
\bibitem [{\citenamefont {Peng}\ \emph {et~al.}(2019)\citenamefont {Peng},
  \citenamefont {Qiao}, \citenamefont {Xian}, \citenamefont {Pan},
  \citenamefont {Ji}, \citenamefont {Zhang},\ and\ \citenamefont
  {Fu}}]{peng2019unusual}%
  \BibitemOpen
  \bibfield  {author} {\bibinfo {author} {\bibfnamefont {L.}~\bibnamefont
  {Peng}}, \bibinfo {author} {\bibfnamefont {J.}~\bibnamefont {Qiao}}, \bibinfo
  {author} {\bibfnamefont {J.-J.}\ \bibnamefont {Xian}}, \bibinfo {author}
  {\bibfnamefont {Y.}~\bibnamefont {Pan}}, \bibinfo {author} {\bibfnamefont
  {W.}~\bibnamefont {Ji}}, \bibinfo {author} {\bibfnamefont {W.}~\bibnamefont
  {Zhang}},\ and\ \bibinfo {author} {\bibfnamefont {Y.-S.}\ \bibnamefont
  {Fu}},\ }\bibfield  {title} {\bibinfo {title} {Unusual electronic states and
  superconducting proximity effect of bi films modulated by a nbse$_2$
  substrate},\ }\href@noop {} {\bibfield  {journal} {\bibinfo  {journal} {ACS
  nano}\ }\textbf {\bibinfo {volume} {13}},\ \bibinfo {pages} {1885} (\bibinfo
  {year} {2019})}\BibitemShut {NoStop}%
\bibitem [{\citenamefont {Yamada}\ \emph {et~al.}(2018)\citenamefont {Yamada},
  \citenamefont {Souma}, \citenamefont {Yamauchi}, \citenamefont {Shimamura},
  \citenamefont {Sugawara}, \citenamefont {Trang}, \citenamefont {Oguchi},
  \citenamefont {Ueno}, \citenamefont {Takahashi},\ and\ \citenamefont
  {Sato}}]{yamada2018ultrathin}%
  \BibitemOpen
  \bibfield  {author} {\bibinfo {author} {\bibfnamefont {K.}~\bibnamefont
  {Yamada}}, \bibinfo {author} {\bibfnamefont {S.}~\bibnamefont {Souma}},
  \bibinfo {author} {\bibfnamefont {K.}~\bibnamefont {Yamauchi}}, \bibinfo
  {author} {\bibfnamefont {N.}~\bibnamefont {Shimamura}}, \bibinfo {author}
  {\bibfnamefont {K.}~\bibnamefont {Sugawara}}, \bibinfo {author}
  {\bibfnamefont {C.~X.}\ \bibnamefont {Trang}}, \bibinfo {author}
  {\bibfnamefont {T.}~\bibnamefont {Oguchi}}, \bibinfo {author} {\bibfnamefont
  {K.}~\bibnamefont {Ueno}}, \bibinfo {author} {\bibfnamefont {T.}~\bibnamefont
  {Takahashi}},\ and\ \bibinfo {author} {\bibfnamefont {T.}~\bibnamefont
  {Sato}},\ }\bibfield  {title} {\bibinfo {title} {Ultrathin bismuth film on
  1t-tas$_2$: Structural transition and charge-density-wave proximity effect},\
  }\href@noop {} {\bibfield  {journal} {\bibinfo  {journal} {Nano letters}\
  }\textbf {\bibinfo {volume} {18}},\ \bibinfo {pages} {3235} (\bibinfo {year}
  {2018})}\BibitemShut {NoStop}%
\bibitem [{\citenamefont {Bian}\ \emph {et~al.}(2009)\citenamefont {Bian},
  \citenamefont {Miller},\ and\ \citenamefont {Chiang}}]{PhysRevB.80.245407}%
  \BibitemOpen
  \bibfield  {author} {\bibinfo {author} {\bibfnamefont {G.}~\bibnamefont
  {Bian}}, \bibinfo {author} {\bibfnamefont {T.}~\bibnamefont {Miller}},\ and\
  \bibinfo {author} {\bibfnamefont {T.-C.}\ \bibnamefont {Chiang}},\ }\bibfield
   {title} {\bibinfo {title} {Electronic structure and surface-mediated
  metastability of bi films on si(111)-$7\ifmmode\times\else\texttimes\fi{}7$
  studied by angle-resolved photoemission spectroscopy},\ }\href
  {https://doi.org/10.1103/PhysRevB.80.245407} {\bibfield  {journal} {\bibinfo
  {journal} {Phys. Rev. B}\ }\textbf {\bibinfo {volume} {80}},\ \bibinfo
  {pages} {245407} (\bibinfo {year} {2009})}\BibitemShut {NoStop}%
\bibitem [{\citenamefont {Shimamura}\ \emph {et~al.}(2018)\citenamefont
  {Shimamura}, \citenamefont {Sugawara}, \citenamefont {Sucharitakul},
  \citenamefont {Souma}, \citenamefont {Iwaya}, \citenamefont {Nakayama},
  \citenamefont {Trang}, \citenamefont {Yamauchi}, \citenamefont {Oguchi},
  \citenamefont {Kudo} \emph {et~al.}}]{shimamura2018ultrathin}%
  \BibitemOpen
  \bibfield  {author} {\bibinfo {author} {\bibfnamefont {N.}~\bibnamefont
  {Shimamura}}, \bibinfo {author} {\bibfnamefont {K.}~\bibnamefont {Sugawara}},
  \bibinfo {author} {\bibfnamefont {S.}~\bibnamefont {Sucharitakul}}, \bibinfo
  {author} {\bibfnamefont {S.}~\bibnamefont {Souma}}, \bibinfo {author}
  {\bibfnamefont {K.}~\bibnamefont {Iwaya}}, \bibinfo {author} {\bibfnamefont
  {K.}~\bibnamefont {Nakayama}}, \bibinfo {author} {\bibfnamefont {C.~X.}\
  \bibnamefont {Trang}}, \bibinfo {author} {\bibfnamefont {K.}~\bibnamefont
  {Yamauchi}}, \bibinfo {author} {\bibfnamefont {T.}~\bibnamefont {Oguchi}},
  \bibinfo {author} {\bibfnamefont {K.}~\bibnamefont {Kudo}}, \emph {et~al.},\
  }\bibfield  {title} {\bibinfo {title} {Ultrathin bismuth film on
  high-temperature cuprate superconductor
  ${\mathrm{bi}}_{2}{\mathrm{sr}}_{2}{\mathrm{cacu}}_{2}{\mathrm{o}}_{8+\ensuremath{\delta}}$
  as a candidate of a topological superconductor},\ }\href@noop {} {\bibfield
  {journal} {\bibinfo  {journal} {ACS nano}\ }\textbf {\bibinfo {volume}
  {12}},\ \bibinfo {pages} {10977} (\bibinfo {year} {2018})}\BibitemShut
  {NoStop}%
\bibitem [{\citenamefont {Drozdov}\ \emph {et~al.}(2014)\citenamefont
  {Drozdov}, \citenamefont {Alexandradinata}, \citenamefont {Jeon},
  \citenamefont {Nadj-Perge}, \citenamefont {Ji}, \citenamefont {Cava},
  \citenamefont {Bernevig},\ and\ \citenamefont {Yazdani}}]{drozdov2014one}%
  \BibitemOpen
  \bibfield  {author} {\bibinfo {author} {\bibfnamefont {I.~K.}\ \bibnamefont
  {Drozdov}}, \bibinfo {author} {\bibfnamefont {A.}~\bibnamefont
  {Alexandradinata}}, \bibinfo {author} {\bibfnamefont {S.}~\bibnamefont
  {Jeon}}, \bibinfo {author} {\bibfnamefont {S.}~\bibnamefont {Nadj-Perge}},
  \bibinfo {author} {\bibfnamefont {H.}~\bibnamefont {Ji}}, \bibinfo {author}
  {\bibfnamefont {R.}~\bibnamefont {Cava}}, \bibinfo {author} {\bibfnamefont
  {B.~A.}\ \bibnamefont {Bernevig}},\ and\ \bibinfo {author} {\bibfnamefont
  {A.}~\bibnamefont {Yazdani}},\ }\bibfield  {title} {\bibinfo {title}
  {One-dimensional topological edge states of bismuth bilayers},\ }\href@noop
  {} {\bibfield  {journal} {\bibinfo  {journal} {Nature Physics}\ }\textbf
  {\bibinfo {volume} {10}},\ \bibinfo {pages} {664} (\bibinfo {year}
  {2014})}\BibitemShut {NoStop}%
\bibitem [{\citenamefont {Kim}\ \emph {et~al.}(2014)\citenamefont {Kim},
  \citenamefont {Jin}, \citenamefont {Park}, \citenamefont {Kim}, \citenamefont
  {Jhi}, \citenamefont {Kim},\ and\ \citenamefont {Yeom}}]{kim2014edge}%
  \BibitemOpen
  \bibfield  {author} {\bibinfo {author} {\bibfnamefont {S.~H.}\ \bibnamefont
  {Kim}}, \bibinfo {author} {\bibfnamefont {K.-H.}\ \bibnamefont {Jin}},
  \bibinfo {author} {\bibfnamefont {J.}~\bibnamefont {Park}}, \bibinfo {author}
  {\bibfnamefont {J.~S.}\ \bibnamefont {Kim}}, \bibinfo {author} {\bibfnamefont
  {S.-H.}\ \bibnamefont {Jhi}}, \bibinfo {author} {\bibfnamefont {T.-H.}\
  \bibnamefont {Kim}},\ and\ \bibinfo {author} {\bibfnamefont {H.~W.}\
  \bibnamefont {Yeom}},\ }\bibfield  {title} {\bibinfo {title} {Edge and
  interfacial states in a two-dimensional topological insulator: Bi(111)
  bilayer on bi$_2$te$_2$se},\ }\href@noop {} {\bibfield  {journal} {\bibinfo
  {journal} {Physical Review B}\ }\textbf {\bibinfo {volume} {89}},\ \bibinfo
  {pages} {155436} (\bibinfo {year} {2014})}\BibitemShut {NoStop}%
\bibitem [{\citenamefont {Koroteev}\ \emph {et~al.}(2008)\citenamefont
  {Koroteev}, \citenamefont {Bihlmayer}, \citenamefont {Chulkov},\ and\
  \citenamefont {Bl\"ugel}}]{PhysRevB.77.045428}%
  \BibitemOpen
  \bibfield  {author} {\bibinfo {author} {\bibfnamefont {Y.~M.}\ \bibnamefont
  {Koroteev}}, \bibinfo {author} {\bibfnamefont {G.}~\bibnamefont {Bihlmayer}},
  \bibinfo {author} {\bibfnamefont {E.~V.}\ \bibnamefont {Chulkov}},\ and\
  \bibinfo {author} {\bibfnamefont {S.}~\bibnamefont {Bl\"ugel}},\ }\bibfield
  {title} {\bibinfo {title} {First-principles investigation of structural and
  electronic properties of ultrathin bi films},\ }\href
  {https://doi.org/10.1103/PhysRevB.77.045428} {\bibfield  {journal} {\bibinfo
  {journal} {Phys. Rev. B}\ }\textbf {\bibinfo {volume} {77}},\ \bibinfo
  {pages} {045428} (\bibinfo {year} {2008})}\BibitemShut {NoStop}%
\bibitem [{\citenamefont {Murakami}(2006)}]{PhysRevLett.97.236805}%
  \BibitemOpen
  \bibfield  {author} {\bibinfo {author} {\bibfnamefont {S.}~\bibnamefont
  {Murakami}},\ }\bibfield  {title} {\bibinfo {title} {Quantum spin hall effect
  and enhanced magnetic response by spin-orbit coupling},\ }\href
  {https://doi.org/10.1103/PhysRevLett.97.236805} {\bibfield  {journal}
  {\bibinfo  {journal} {Phys. Rev. Lett.}\ }\textbf {\bibinfo {volume} {97}},\
  \bibinfo {pages} {236805} (\bibinfo {year} {2006})}\BibitemShut {NoStop}%
\bibitem [{\citenamefont {Peng}\ \emph {et~al.}(2018)\citenamefont {Peng},
  \citenamefont {Xian}, \citenamefont {Tang}, \citenamefont {Rubio},
  \citenamefont {Zhang}, \citenamefont {Zhang},\ and\ \citenamefont
  {Fu}}]{peng2018visualizing}%
  \BibitemOpen
  \bibfield  {author} {\bibinfo {author} {\bibfnamefont {L.}~\bibnamefont
  {Peng}}, \bibinfo {author} {\bibfnamefont {J.-J.}\ \bibnamefont {Xian}},
  \bibinfo {author} {\bibfnamefont {P.}~\bibnamefont {Tang}}, \bibinfo {author}
  {\bibfnamefont {A.}~\bibnamefont {Rubio}}, \bibinfo {author} {\bibfnamefont
  {S.-C.}\ \bibnamefont {Zhang}}, \bibinfo {author} {\bibfnamefont
  {W.}~\bibnamefont {Zhang}},\ and\ \bibinfo {author} {\bibfnamefont {Y.-S.}\
  \bibnamefont {Fu}},\ }\bibfield  {title} {\bibinfo {title} {Visualizing
  topological edge states of single and double bilayer bi supported on
  multibilayer bi(111) films},\ }\href@noop {} {\bibfield  {journal} {\bibinfo
  {journal} {Phys. Rev. B}\ }\textbf {\bibinfo {volume} {98}},\ \bibinfo
  {pages} {245108} (\bibinfo {year} {2018})}\BibitemShut {NoStop}%
\bibitem [{\citenamefont {Lu}\ \emph {et~al.}(2015)\citenamefont {Lu},
  \citenamefont {Xu}, \citenamefont {Zeng}, \citenamefont {Yao}, \citenamefont
  {Shen}, \citenamefont {Yang}, \citenamefont {Luo}, \citenamefont {Pan},
  \citenamefont {Wu}, \citenamefont {Das} \emph {et~al.}}]{lu2015topological}%
  \BibitemOpen
  \bibfield  {author} {\bibinfo {author} {\bibfnamefont {Y.}~\bibnamefont
  {Lu}}, \bibinfo {author} {\bibfnamefont {W.}~\bibnamefont {Xu}}, \bibinfo
  {author} {\bibfnamefont {M.}~\bibnamefont {Zeng}}, \bibinfo {author}
  {\bibfnamefont {G.}~\bibnamefont {Yao}}, \bibinfo {author} {\bibfnamefont
  {L.}~\bibnamefont {Shen}}, \bibinfo {author} {\bibfnamefont {M.}~\bibnamefont
  {Yang}}, \bibinfo {author} {\bibfnamefont {Z.}~\bibnamefont {Luo}}, \bibinfo
  {author} {\bibfnamefont {F.}~\bibnamefont {Pan}}, \bibinfo {author}
  {\bibfnamefont {K.}~\bibnamefont {Wu}}, \bibinfo {author} {\bibfnamefont
  {T.}~\bibnamefont {Das}}, \emph {et~al.},\ }\bibfield  {title} {\bibinfo
  {title} {Topological properties determined by atomic buckling in
  self-assembled ultrathin bi(110)},\ }\href@noop {} {\bibfield  {journal}
  {\bibinfo  {journal} {Nano letters}\ }\textbf {\bibinfo {volume} {15}},\
  \bibinfo {pages} {80} (\bibinfo {year} {2015})}\BibitemShut {NoStop}%
\bibitem [{\citenamefont {Yang}\ \emph {et~al.}(2017)\citenamefont {Yang},
  \citenamefont {Jandke}, \citenamefont {Storbeck}, \citenamefont {Balashov},
  \citenamefont {Aishwarya},\ and\ \citenamefont {Wulfhekel}}]{yang2017edge}%
  \BibitemOpen
  \bibfield  {author} {\bibinfo {author} {\bibfnamefont {F.}~\bibnamefont
  {Yang}}, \bibinfo {author} {\bibfnamefont {J.}~\bibnamefont {Jandke}},
  \bibinfo {author} {\bibfnamefont {T.}~\bibnamefont {Storbeck}}, \bibinfo
  {author} {\bibfnamefont {T.}~\bibnamefont {Balashov}}, \bibinfo {author}
  {\bibfnamefont {A.}~\bibnamefont {Aishwarya}},\ and\ \bibinfo {author}
  {\bibfnamefont {W.}~\bibnamefont {Wulfhekel}},\ }\bibfield  {title} {\bibinfo
  {title} {Edge states in mesoscopic bi islands on superconducting nb(110)},\
  }\href@noop {} {\bibfield  {journal} {\bibinfo  {journal} {Physical Review
  B}\ }\textbf {\bibinfo {volume} {96}},\ \bibinfo {pages} {235413} (\bibinfo
  {year} {2017})}\BibitemShut {NoStop}%
\bibitem [{\citenamefont {Gou}\ \emph {et~al.}(2020)\citenamefont {Gou},
  \citenamefont {Kong}, \citenamefont {He}, \citenamefont {Huang},
  \citenamefont {Sun}, \citenamefont {Meng}, \citenamefont {Wu}, \citenamefont
  {Chen},\ and\ \citenamefont {Wee}}]{gou2020effect}%
  \BibitemOpen
  \bibfield  {author} {\bibinfo {author} {\bibfnamefont {J.}~\bibnamefont
  {Gou}}, \bibinfo {author} {\bibfnamefont {L.}~\bibnamefont {Kong}}, \bibinfo
  {author} {\bibfnamefont {X.}~\bibnamefont {He}}, \bibinfo {author}
  {\bibfnamefont {Y.~L.}\ \bibnamefont {Huang}}, \bibinfo {author}
  {\bibfnamefont {J.}~\bibnamefont {Sun}}, \bibinfo {author} {\bibfnamefont
  {S.}~\bibnamefont {Meng}}, \bibinfo {author} {\bibfnamefont {K.}~\bibnamefont
  {Wu}}, \bibinfo {author} {\bibfnamefont {L.}~\bibnamefont {Chen}},\ and\
  \bibinfo {author} {\bibfnamefont {A.~T.~S.}\ \bibnamefont {Wee}},\ }\bibfield
   {title} {\bibinfo {title} {The effect of moir{\'e} superstructures on
  topological edge states in twisted bismuthene homojunctions},\ }\href@noop {}
  {\bibfield  {journal} {\bibinfo  {journal} {Science Advances}\ }\textbf
  {\bibinfo {volume} {6}},\ \bibinfo {pages} {eaba2773} (\bibinfo {year}
  {2020})}\BibitemShut {NoStop}%
\bibitem [{\citenamefont {Sun}\ \emph {et~al.}(2012)\citenamefont {Sun},
  \citenamefont {Huang}, \citenamefont {Wong}, \citenamefont {Gao},
  \citenamefont {Feng},\ and\ \citenamefont {Wee}}]{sun2012energy}%
  \BibitemOpen
  \bibfield  {author} {\bibinfo {author} {\bibfnamefont {J.-T.}\ \bibnamefont
  {Sun}}, \bibinfo {author} {\bibfnamefont {H.}~\bibnamefont {Huang}}, \bibinfo
  {author} {\bibfnamefont {S.~L.}\ \bibnamefont {Wong}}, \bibinfo {author}
  {\bibfnamefont {H.-J.}\ \bibnamefont {Gao}}, \bibinfo {author} {\bibfnamefont
  {Y.~P.}\ \bibnamefont {Feng}},\ and\ \bibinfo {author} {\bibfnamefont
  {A.~T.~S.}\ \bibnamefont {Wee}},\ }\bibfield  {title} {\bibinfo {title}
  {Energy-gap opening in a bi(110) nanoribbon induced by edge reconstruction},\
  }\href@noop {} {\bibfield  {journal} {\bibinfo  {journal} {Physical review
  letters}\ }\textbf {\bibinfo {volume} {109}},\ \bibinfo {pages} {246804}
  (\bibinfo {year} {2012})}\BibitemShut {NoStop}%
\bibitem [{\citenamefont {Yaginuma}\ \emph {et~al.}(2007)\citenamefont
  {Yaginuma}, \citenamefont {Nagaoka}, \citenamefont {Nagao}, \citenamefont
  {Bihlmayer}, \citenamefont {M.~Koroteev}, \citenamefont {V.~Chulkov},\ and\
  \citenamefont {Nakayama}}]{yaginuma2007electronic}%
  \BibitemOpen
  \bibfield  {author} {\bibinfo {author} {\bibfnamefont {S.}~\bibnamefont
  {Yaginuma}}, \bibinfo {author} {\bibfnamefont {K.}~\bibnamefont {Nagaoka}},
  \bibinfo {author} {\bibfnamefont {T.}~\bibnamefont {Nagao}}, \bibinfo
  {author} {\bibfnamefont {G.}~\bibnamefont {Bihlmayer}}, \bibinfo {author}
  {\bibfnamefont {Y.}~\bibnamefont {M.~Koroteev}}, \bibinfo {author}
  {\bibfnamefont {E.}~\bibnamefont {V.~Chulkov}},\ and\ \bibinfo {author}
  {\bibfnamefont {T.}~\bibnamefont {Nakayama}},\ }\bibfield  {title} {\bibinfo
  {title} {Electronic structure of ultrathin bismuth films with a7 and
  black-phosphorus-like structures},\ }\href@noop {} {\bibfield  {journal}
  {\bibinfo  {journal} {Journal of the Physical Society of Japan}\ }\textbf
  {\bibinfo {volume} {77}},\ \bibinfo {pages} {014701} (\bibinfo {year}
  {2007})}\BibitemShut {NoStop}%
\bibitem [{\citenamefont {Kokubo}\ \emph {et~al.}(2015)\citenamefont {Kokubo},
  \citenamefont {Yoshiike}, \citenamefont {Nakatsuji},\ and\ \citenamefont
  {Hirayama}}]{kokubo2015ultrathin}%
  \BibitemOpen
  \bibfield  {author} {\bibinfo {author} {\bibfnamefont {I.}~\bibnamefont
  {Kokubo}}, \bibinfo {author} {\bibfnamefont {Y.}~\bibnamefont {Yoshiike}},
  \bibinfo {author} {\bibfnamefont {K.}~\bibnamefont {Nakatsuji}},\ and\
  \bibinfo {author} {\bibfnamefont {H.}~\bibnamefont {Hirayama}},\ }\bibfield
  {title} {\bibinfo {title} {Ultrathin bi(110) films on si(111)3$\times$ 3-b
  substrates},\ }\href@noop {} {\bibfield  {journal} {\bibinfo  {journal}
  {Phys. Rev. B}\ }\textbf {\bibinfo {volume} {91}},\ \bibinfo {pages} {075429}
  (\bibinfo {year} {2015})}\BibitemShut {NoStop}%
\bibitem [{\citenamefont {Kowalczyk}\ \emph {et~al.}(2020)\citenamefont
  {Kowalczyk}, \citenamefont {Brown}, \citenamefont {Maerkl}, \citenamefont
  {Lu}, \citenamefont {Chiu}, \citenamefont {Liu}, \citenamefont {Yang},
  \citenamefont {Wang}, \citenamefont {Zasada}, \citenamefont {Genuzio} \emph
  {et~al.}}]{kowalczyk2020realization}%
  \BibitemOpen
  \bibfield  {author} {\bibinfo {author} {\bibfnamefont {P.~J.}\ \bibnamefont
  {Kowalczyk}}, \bibinfo {author} {\bibfnamefont {S.~A.}\ \bibnamefont
  {Brown}}, \bibinfo {author} {\bibfnamefont {T.}~\bibnamefont {Maerkl}},
  \bibinfo {author} {\bibfnamefont {Q.}~\bibnamefont {Lu}}, \bibinfo {author}
  {\bibfnamefont {C.-K.}\ \bibnamefont {Chiu}}, \bibinfo {author}
  {\bibfnamefont {Y.}~\bibnamefont {Liu}}, \bibinfo {author} {\bibfnamefont
  {S.~A.}\ \bibnamefont {Yang}}, \bibinfo {author} {\bibfnamefont
  {X.}~\bibnamefont {Wang}}, \bibinfo {author} {\bibfnamefont {I.}~\bibnamefont
  {Zasada}}, \bibinfo {author} {\bibfnamefont {F.}~\bibnamefont {Genuzio}},
  \emph {et~al.},\ }\bibfield  {title} {\bibinfo {title} {Realization of
  symmetry-enforced two-dimensional dirac fermions in nonsymmorphic
  $\alpha$-bismuthene},\ }\href@noop {} {\bibfield  {journal} {\bibinfo
  {journal} {ACS nano}\ }\textbf {\bibinfo {volume} {14}},\ \bibinfo {pages}
  {1888} (\bibinfo {year} {2020})}\BibitemShut {NoStop}%
\bibitem [{\citenamefont {Hatta}\ \emph {et~al.}(2009)\citenamefont {Hatta},
  \citenamefont {Ohtsubo}, \citenamefont {Miyamoto}, \citenamefont {Okuyama},\
  and\ \citenamefont {Aruga}}]{hatta2009epitaxial}%
  \BibitemOpen
  \bibfield  {author} {\bibinfo {author} {\bibfnamefont {S.}~\bibnamefont
  {Hatta}}, \bibinfo {author} {\bibfnamefont {Y.}~\bibnamefont {Ohtsubo}},
  \bibinfo {author} {\bibfnamefont {S.}~\bibnamefont {Miyamoto}}, \bibinfo
  {author} {\bibfnamefont {H.}~\bibnamefont {Okuyama}},\ and\ \bibinfo {author}
  {\bibfnamefont {T.}~\bibnamefont {Aruga}},\ }\bibfield  {title} {\bibinfo
  {title} {Epitaxial growth of bi thin films on ge (111)},\ }\href@noop {}
  {\bibfield  {journal} {\bibinfo  {journal} {Applied Surface Science}\
  }\textbf {\bibinfo {volume} {256}},\ \bibinfo {pages} {1252} (\bibinfo {year}
  {2009})}\BibitemShut {NoStop}%
\bibitem [{\citenamefont {Lindberg}\ \emph {et~al.}(1988)\citenamefont
  {Lindberg}, \citenamefont {Shen}, \citenamefont {Wells}, \citenamefont
  {Mitzi}, \citenamefont {Lindau}, \citenamefont {Spicer},\ and\ \citenamefont
  {Kapitulnik}}]{lindberg1988surface}%
  \BibitemOpen
  \bibfield  {author} {\bibinfo {author} {\bibfnamefont {P.}~\bibnamefont
  {Lindberg}}, \bibinfo {author} {\bibfnamefont {Z.-X.}\ \bibnamefont {Shen}},
  \bibinfo {author} {\bibfnamefont {B.}~\bibnamefont {Wells}}, \bibinfo
  {author} {\bibfnamefont {D.}~\bibnamefont {Mitzi}}, \bibinfo {author}
  {\bibfnamefont {I.}~\bibnamefont {Lindau}}, \bibinfo {author} {\bibfnamefont
  {W.}~\bibnamefont {Spicer}},\ and\ \bibinfo {author} {\bibfnamefont
  {A.}~\bibnamefont {Kapitulnik}},\ }\bibfield  {title} {\bibinfo {title}
  {Surface structure of
  ${\mathrm{bi}}_{2}{\mathrm{sr}}_{2}{\mathrm{cacu}}_{2}{\mathrm{o}}_{8+\ensuremath{\delta}}$
  high-temperature superconductors studied using low-energy electron
  diffraction},\ }\href@noop {} {\bibfield  {journal} {\bibinfo  {journal}
  {Applied physics letters}\ }\textbf {\bibinfo {volume} {53}},\ \bibinfo
  {pages} {2563} (\bibinfo {year} {1988})}\BibitemShut {NoStop}%
\bibitem [{\citenamefont {Drozdov}\ \emph {et~al.}(2018)\citenamefont
  {Drozdov}, \citenamefont {Pletikosi{\'c}}, \citenamefont {Kim}, \citenamefont
  {Fujita}, \citenamefont {Gu}, \citenamefont {Davis}, \citenamefont {Johnson},
  \citenamefont {Bo{\v{z}}ovi{\'c}},\ and\ \citenamefont
  {Valla}}]{drozdov2018}%
  \BibitemOpen
  \bibfield  {author} {\bibinfo {author} {\bibfnamefont {I.~K.}\ \bibnamefont
  {Drozdov}}, \bibinfo {author} {\bibfnamefont {I.}~\bibnamefont
  {Pletikosi{\'c}}}, \bibinfo {author} {\bibfnamefont {C.-K.}\ \bibnamefont
  {Kim}}, \bibinfo {author} {\bibfnamefont {K.}~\bibnamefont {Fujita}},
  \bibinfo {author} {\bibfnamefont {G.}~\bibnamefont {Gu}}, \bibinfo {author}
  {\bibfnamefont {J.~S.}\ \bibnamefont {Davis}}, \bibinfo {author}
  {\bibfnamefont {P.}~\bibnamefont {Johnson}}, \bibinfo {author} {\bibfnamefont
  {I.}~\bibnamefont {Bo{\v{z}}ovi{\'c}}},\ and\ \bibinfo {author}
  {\bibfnamefont {T.}~\bibnamefont {Valla}},\ }\bibfield  {title} {\bibinfo
  {title} {Phase diagram of bi 2 sr 2 cacu 2 o 8+ $\delta$ revisited},\
  }\href@noop {} {\bibfield  {journal} {\bibinfo  {journal} {Nature
  communications}\ }\textbf {\bibinfo {volume} {9}},\ \bibinfo {pages} {1}
  (\bibinfo {year} {2018})}\BibitemShut {NoStop}%
\bibitem [{\citenamefont {Valla}\ \emph {et~al.}(2020)\citenamefont {Valla},
  \citenamefont {Drozdov},\ and\ \citenamefont {Gu}}]{Valla2020}%
  \BibitemOpen
  \bibfield  {author} {\bibinfo {author} {\bibfnamefont {T.}~\bibnamefont
  {Valla}}, \bibinfo {author} {\bibfnamefont {I.~K.}\ \bibnamefont {Drozdov}},\
  and\ \bibinfo {author} {\bibfnamefont {G.~D.}\ \bibnamefont {Gu}},\
  }\bibfield  {title} {\bibinfo {title} {{Disappearance of Superconductivity
  Due to Vanishing Coupling in the Overdoped
  ${\mathrm{Bi}}_{2}{\mathrm{Sr}}_{2}{\mathrm{CaCu}}_{2}{\mathrm{O}}_{8+\ensuremath{\delta}}$}},\
  }\href@noop {} {\bibfield  {journal} {\bibinfo  {journal} {Nature
  Communications}\ }\textbf {\bibinfo {volume} {11}},\ \bibinfo {pages} {569}
  (\bibinfo {year} {2020})}\BibitemShut {NoStop}%
\bibitem [{\citenamefont {Kowalczyk}\ \emph {et~al.}(2013)\citenamefont
  {Kowalczyk}, \citenamefont {Mahapatra}, \citenamefont {Brown}, \citenamefont
  {Bian}, \citenamefont {Wang},\ and\ \citenamefont {Chiang}}]{kowalczyk2013}%
  \BibitemOpen
  \bibfield  {author} {\bibinfo {author} {\bibfnamefont {P.~J.}\ \bibnamefont
  {Kowalczyk}}, \bibinfo {author} {\bibfnamefont {O.}~\bibnamefont
  {Mahapatra}}, \bibinfo {author} {\bibfnamefont {S.}~\bibnamefont {Brown}},
  \bibinfo {author} {\bibfnamefont {G.}~\bibnamefont {Bian}}, \bibinfo {author}
  {\bibfnamefont {X.}~\bibnamefont {Wang}},\ and\ \bibinfo {author}
  {\bibfnamefont {T.-C.}\ \bibnamefont {Chiang}},\ }\bibfield  {title}
  {\bibinfo {title} {Electronic size effects in three-dimensional
  nanostructures},\ }\href@noop {} {\bibfield  {journal} {\bibinfo  {journal}
  {Nano letters}\ }\textbf {\bibinfo {volume} {13}},\ \bibinfo {pages} {43}
  (\bibinfo {year} {2013})}\BibitemShut {NoStop}%
\bibitem [{\citenamefont {Agergaard}\ \emph {et~al.}(2001)\citenamefont
  {Agergaard}, \citenamefont {S{\o}ndergaard}, \citenamefont {Li},
  \citenamefont {Nielsen}, \citenamefont {Hoffmann}, \citenamefont {Li},\ and\
  \citenamefont {Hofmann}}]{agergaard2001effect}%
  \BibitemOpen
  \bibfield  {author} {\bibinfo {author} {\bibfnamefont {S.}~\bibnamefont
  {Agergaard}}, \bibinfo {author} {\bibfnamefont {C.}~\bibnamefont
  {S{\o}ndergaard}}, \bibinfo {author} {\bibfnamefont {H.}~\bibnamefont {Li}},
  \bibinfo {author} {\bibfnamefont {M.~B.}\ \bibnamefont {Nielsen}}, \bibinfo
  {author} {\bibfnamefont {S.}~\bibnamefont {Hoffmann}}, \bibinfo {author}
  {\bibfnamefont {Z.}~\bibnamefont {Li}},\ and\ \bibinfo {author}
  {\bibfnamefont {P.}~\bibnamefont {Hofmann}},\ }\bibfield  {title} {\bibinfo
  {title} {The effect of reduced dimensionality on a semimetal: the electronic
  structure of the bi(110) surface},\ }\href@noop {} {\bibfield  {journal}
  {\bibinfo  {journal} {New journal of physics}\ }\textbf {\bibinfo {volume}
  {3}},\ \bibinfo {pages} {15} (\bibinfo {year} {2001})}\BibitemShut {NoStop}%
\bibitem [{\citenamefont {Kundu}\ \emph {et~al.}(2020)\citenamefont {Kundu},
  \citenamefont {Wu}, \citenamefont {Drozdov}, \citenamefont {Gu},\ and\
  \citenamefont {Valla}}]{kundu2020origin}%
  \BibitemOpen
  \bibfield  {author} {\bibinfo {author} {\bibfnamefont {A.~K.}\ \bibnamefont
  {Kundu}}, \bibinfo {author} {\bibfnamefont {Z.-B.}\ \bibnamefont {Wu}},
  \bibinfo {author} {\bibfnamefont {I.}~\bibnamefont {Drozdov}}, \bibinfo
  {author} {\bibfnamefont {G.}~\bibnamefont {Gu}},\ and\ \bibinfo {author}
  {\bibfnamefont {T.}~\bibnamefont {Valla}},\ }\bibfield  {title} {\bibinfo
  {title} {Origin of suppression of proximity-induced superconductivity in
  bi/${\mathrm{bi}}_{2}{\mathrm{sr}}_{2}{\mathrm{cacu}}_{2}{\mathrm{o}}_{8+\ensuremath{\delta}}$
  heterostructures},\ }\href@noop {} {\bibfield  {journal} {\bibinfo  {journal}
  {Advanced Quantum Technologies}\ }\textbf {\bibinfo {volume} {3}},\ \bibinfo
  {pages} {2000038} (\bibinfo {year} {2020})}\BibitemShut {NoStop}%
\bibitem [{\citenamefont {Yilmaz}\ \emph {et~al.}(2014)\citenamefont {Yilmaz},
  \citenamefont {Pletikosi\ifmmode~\acute{c}\else \'{c}\fi{}}, \citenamefont
  {Weber}, \citenamefont {Sadowski}, \citenamefont {Gu}, \citenamefont
  {Caruso}, \citenamefont {Sinkovic},\ and\ \citenamefont
  {Valla}}]{yilmaz2014absence}%
  \BibitemOpen
  \bibfield  {author} {\bibinfo {author} {\bibfnamefont {T.}~\bibnamefont
  {Yilmaz}}, \bibinfo {author} {\bibfnamefont {I.}~\bibnamefont
  {Pletikosi\ifmmode~\acute{c}\else \'{c}\fi{}}}, \bibinfo {author}
  {\bibfnamefont {A.~P.}\ \bibnamefont {Weber}}, \bibinfo {author}
  {\bibfnamefont {J.~T.}\ \bibnamefont {Sadowski}}, \bibinfo {author}
  {\bibfnamefont {G.~D.}\ \bibnamefont {Gu}}, \bibinfo {author} {\bibfnamefont
  {A.~N.}\ \bibnamefont {Caruso}}, \bibinfo {author} {\bibfnamefont
  {B.}~\bibnamefont {Sinkovic}},\ and\ \bibinfo {author} {\bibfnamefont
  {T.}~\bibnamefont {Valla}},\ }\bibfield  {title} {\bibinfo {title} {Absence
  of a proximity effect for a thin-films of a
  ${\mathrm{bi}}_{2}{\mathrm{se}}_{3}$ topological insulator grown on top of a
  ${\mathrm{bi}}_{2}{\mathrm{sr}}_{2}{\mathrm{cacu}}_{2}{\mathrm{o}}_{8+\ensuremath{\delta}}$
  cuprate superconductor},\ }\href
  {https://doi.org/10.1103/PhysRevLett.113.067003} {\bibfield  {journal}
  {\bibinfo  {journal} {Phys. Rev. Lett.}\ }\textbf {\bibinfo {volume} {113}},\
  \bibinfo {pages} {067003} (\bibinfo {year} {2014})}\BibitemShut {NoStop}%
\bibitem [{\citenamefont {Kim}\ \emph {et~al.}(2018)\citenamefont {Kim},
  \citenamefont {Drozdov}, \citenamefont {Fujita}, \citenamefont {Davis},
  \citenamefont {Bo{\v{z}}ovi{\'c}},\ and\ \citenamefont {Valla}}]{Kim2018a}%
  \BibitemOpen
  \bibfield  {author} {\bibinfo {author} {\bibfnamefont {C.~K.}\ \bibnamefont
  {Kim}}, \bibinfo {author} {\bibfnamefont {I.~K.}\ \bibnamefont {Drozdov}},
  \bibinfo {author} {\bibfnamefont {K.}~\bibnamefont {Fujita}}, \bibinfo
  {author} {\bibfnamefont {J.~S.}\ \bibnamefont {Davis}}, \bibinfo {author}
  {\bibfnamefont {I.}~\bibnamefont {Bo{\v{z}}ovi{\'c}}},\ and\ \bibinfo
  {author} {\bibfnamefont {T.}~\bibnamefont {Valla}},\ }\bibfield  {title}
  {\bibinfo {title} {In-situ angle-resolved photoemission spectroscopy of
  copper-oxide thin films synthesized by molecular beam epitaxy},\ }\bibfield
  {journal} {\bibinfo  {journal} {Journal of Electron Spectroscopy and Related
  Phenomena}\ }\href
  {https://doi.org/https://doi.org/10.1016/j.elspec.2018.07.003}
  {https://doi.org/10.1016/j.elspec.2018.07.003} (\bibinfo {year}
  {2018})\BibitemShut {NoStop}%
\end{thebibliography}%

\end{document}